\documentclass[12pt]{article}

\usepackage{graphicx}
\usepackage{color}

\begin{document}

\def\beq{\begin{equation}}
\def\eeq{\end{equation}}
\newcommand{\comment}[1]{}

\begin{center}
{\Large \bf \boldmath
Use of $Z$ polarization in $e^+e^- \to ZH$ to measure the triple-Higgs
coupling}
\end{center}

\bigskip\smallskip

\begin{center}
Kumar Rao$^a$, Saurabh D. Rindani$^b$, Priyanka Sarmah$^a$, Balbeer
Singh$^c$

\bigskip\smallskip
{\it $^a$Department of Physics, Indian Institute of Technology,\\ Powai,
Mumbai 400076, India}

\smallskip

{\it $^b$Theoretical Physics Division, Physical Research Laboratory,\\
Navrangpura, Ahmedabad 380009, India
}

\smallskip

{\it $^c$Department of Theoretical Physics,\\ Tata Institute of
Fundamental Research,
Dr. Homi Bhabha Road,\\ Colaba, Mumbai 400005, India}

\bigskip \smallskip

{\small \bf Abstract}

\end{center}
\smallskip

It is shown that a certain angular asymmetry of charged leptons produced in the decay of $Z$ in the
process $e^+e^- \to ZH$, related to a tensor polarization
component of the $Z$, can be used to constrain
the anomalous triple-Higgs coupling,
independent of the other anomalous couplings like the $ZZH$ coupling
which dominate at tree level. 
This is because the angular asymmetry is odd under na\" ive time
reversal, and hence dependent on loop-level contributions.
At a future $e^+e^-$ collider like the
International Linear Collider (ILC), for example, a limit of about 
$3.4$ might be placed on the ratio of the actual triple Higgs coupling 
to that predicted in the standard model for a centre-of-mass energy of
500 GeV and an integrated luminosity of 30 fb$^{-1}$ with electron and
positron beams having longitudinal polarization of $-80\%$ and $+30\%$,
respectively.

\section{Anomalous triple-Higgs coupling}

Since the discovery of the Higgs boson, all-out efforts are being made
to determine the properties of the Higgs boson, especially its couplings to
fermions and gauge bosons, with increasing accuracy. The results seem to
be in
agreement with the predictions of the standard model (SM) 
to a good degree
of accuracy. An important aspect to pin down the theory as 
the SM is to measure the
scalar self-couplings with great accuracy and to check their agreement 
with the predictions of the SM.

The scalar self-couplings correspond to $\lambda_3$ and
$\lambda_4$ in the following terms in the scalar Lagrangian
\beq\label{3H}
{\cal L}_{3H} = - \lambda_3 H^3,
\eeq
\beq\label{4H}
{\cal L}_{4H} = - \lambda_4 H^4.
\eeq
In the SM, these couplings are related to the physical Higgs mass $m_H$
and the scalar vacuum expectation value $v$ by
\beq\label{SMlambdas}
\lambda_3^{\mathrm SM} = \lambda_4^{\mathrm SM} v ; \;\;
\lambda_4^{\mathrm SM} = \frac{m_H^2}{2 v^2}.
\eeq

Future experiments at the LHC as well as at the
proposed lepton colliders will determine 
$\lambda_3$ and $\lambda_4$ with
greater precision and would be able to check the SM relations of 
eq. (\ref{SMlambdas}).
It is possible that the correct full theory is not the SM, but an
extension of the SM. In that case, the above couplings could be
the couplings in an effective theory, and they may not obey the
relations (\ref{SMlambdas}). The deviations of these couplings from
their SM values have been discussed in the context of the standard model
effective field theory (SMEFT), 
where effective interactions induced by new physics
are written in terms of higher-dimensional operators, suppressed by a
high-energy scale, the effective theory presumed to be valid at energies
much lower than this scale. Thus, for example, ${\cal L}_{3H}$ would get
a contribution from a dimension-six operator, $-\lambda_6 (H^\dag H)^3$,
see for example, \cite{Degrassi:2016wml}.

A determination of the triple-Higgs coupling $\lambda_3$ can be
carried out through a process where two (or more) Higgs bosons are
produced. First of all, such a process needs a high centre-of-mass
(c.m.) energy.
Moreover, it has been found that in the SM \cite{glover}, there is
destructive interference between the one-loop diagrams contributing 
to the process $ gg \to HH$,
making the total cross section extremely small at a hadron collider. 
Thus, the accuracy of
the determination of $\lambda_3$ is low.

A suggestion was made by McCullough \cite{McCullough:2013rea} 
that the triple-Higgs
coupling could be measured through its contribution in one-loop diagrams
in single-Higgs production. 
The process considered in \cite{McCullough:2013rea}
was $e^+e^- \to ZH$, in which, it was assumed that only
$\lambda_3$ deviates from its SM value,
\beq\label{kappa}
\lambda_3 = \lambda_3^{\mathrm SM} (1 + \kappa).
\eeq
$\kappa$ can be nonzero in an SMEFT, for example,
in the presence of dimension-six $\phi^6$ operators. 
The conclusion was that it would be possible to put a limit on the 
fractional deviation $\kappa$
which is of the order of 28\% at $e^+e^-$ c.m. energy of 240 GeV, with
an integrated luminosity ${\cal L}$ of 10 ab$^{-1}$, expected to be 
available at
TLEP (currently known as FCC-ee) \cite{TLEP}.
It was shown \cite{Rindani:2018ubx}
that the sensitivity of
$e^+e^- \to ZH$ can be improved with polarized beams 
and, in particular, for the polarization combination 
($P_{e^-} = -0.8$, $P_{e^+} =
+0.3$) the accuracy to measure $\kappa$ is about 57\% for
$\sqrt{s}=250$ GeV and ${\cal L}=2$ ab$^{-1}$,
envisaged at the ILC, as compared to 70\% for the case of {unpolarized }
beams. 
This estimate is based on the assumption, made
for the sake of concreteness, that there are no other contributions 
to an effective $ZZH$ vertex. 

Unfortunately, anomalous $ZZH$ couplings, as for example, characterized
by dimension-six operators in SMEFT, 
can contribute at tree level to the cross section for $e^+e^- \to ZH$,
and hence this contribution can overwhelm the triple-Higgs contribution.
So this sensitivity is possible only provided the tree-level 
contributions are known, or eliminated somehow.
The extraction of $\kappa$
from the cross section would then be possible, for example, 
making use of
measurement at more than one energy, using the
different energy dependences of the two contributions
\cite{McCullough:2013rea}, or using details of $Z$ decay distributions.

We examine here the possibility that $Z$ polarization in $e^+e^- \to ZH$
can be used to measure $\kappa$ with either less sensitivity to the
tree-level $ZZH$ coupling, or, perhaps, independent of it.
Using a very useful and simple relation first formulated in
\cite{Boudjema:2009fz}, the polarization parameters of the $Z$ produced 
in various processes can be accessed through angular
asymmetries of the decay products of the $Z$, especially a charged
lepton pair \cite{Boudjema:2009fz, Rahaman:2017qql, Rahaman:2016pqj,
Rao:2019hsp, Rao:2020hel}. 
This property was applied specifically to the process $e^+e^- \to ZH$
in \cite{Rao:2019hsp}.
The interesting feature of $Z$ polarization and the consequent decay 
asymmetry that we would like to exploit here is that one particular
asymmetry gets contribution from only the absorptive part of the
amplitude. 
It is therefore sensitive to the loop-level
contribution of the triple-Higgs coupling. This particular 
CP-even angular asymmetry is odd under time 
reversal T\footnote{ 
We refer here to na\" ive time reversal, that is, a reversal of the  
directions of all 
spins and momenta, but not an interchange of initial and final states as
is required by genuine time reversal. Henceforth T will refer to na\"
ive time reversal}.
The CPT theorem then requires 
the existence of an absorptive part in the amplitude for this asymmetry
to be nonzero. This asymmetry then automatically
measures the interference between the tree amplitude and the 
loop amplitude. Since the tree-level $ZZH$ coupling at the lowest order
is the SM coupling, the asymmetry is a measure of the loop contribution
when restricting  to linear order in anomalous couplings\footnote{The 
fact that na\" ive T-odd quantities can be
used for studying loop induced triple-Higgs coupling has been made use
of in \cite{Nakamura:2018bli}, though they do not relate it to $Z$
polarization variables.}. 

While it is true that this asymmetry gets
contribution from whichever loop amplitude contributes to the process,
we are mainly interested in the triple-Higgs coupling which is likely to
be only mildly constrained. As compared to other couplings entering loop
amplitudes, like the top Yukawa coupling or the $WWH$ coupling, which
would already be well known and close to their SM values, the
triple-Higgs coupling would contribute the dominant uncertainty.

There have been various suggestions for the
construction of $e^+e^-$ colliders with c.m. energy ranging from a few
hundred GeV to a few TeV. After the discovery of the Higgs boson with
mass of about 125 GeV, the dominant suggestion is to construct a linear
collider, named the International Linear Collider, which would
first operate at a c.m. energy of 250 GeV, enabling precise measurement
of Higgs properties, through an abundant production of a $ZH$ final 
state.  There have also been other proposals, as for example, the 
Compact Linear
Collider (CLIC), the Future Circular Collider (FCC-ee) and Circular
Electron Positron Collider (CEPC) where electron and positron beams
would be collided, providing a clean environment to study couplings of
SM particles, and possibly look for new physics, if any. 
In the context
of $e^+e^-$ colliders, particularly the ILC, the possibility of
utilizing beam polarization and its advantages has received much
attention. Suitable longitudinal beam polarization could help 
in improving the sensitivity for
many different processes and suppressing unwanted background 
\cite{MoortgatPick:2005cw, Fujii:2018mli}.

The use of polarization in the context of Higgs properties is also
discussed in \cite{Durieux:2017rsg, Barklow:2017suo}. 
It is expected that at the ILC, polarizations of 80\% and 30\% would be
possible respectively for electron and positron beams for c.m. energy of
250 GeV \cite{Fujii:2018mli}.

On the experimental side, polarization of weak gauge bosons has been 
measured at the LHC in $W + \rm jet$ production
\cite{CMS:2011kaj,
ATLAS:2012au},
$Z + \rm jet$ production 
\cite{CMS:2015cyj, ATLAS:2016rnf},
in $W$ production in the decay of top quarks 
\cite{helicityLHC}, 
and more recently in $WZ$ production
\cite{ATLAS:2019bsc}
and same-sign $WW$ production
\cite{CMS:2020etf}.
The gauge-boson polarizations and helicity fractions are 
inferred from the angular distributions of the fermions to which the
gauge bosons decay \cite{Stirling:2012zt}.

Various methods have been suggested for investigating 
the Higgs self-coupling. In a recent study \cite{Amacker:2020bmn}, it is
shown how the  
trilinear Higgs coupling can be  constrained from di-Higgs production 
with a $b\bar{b}b\bar{b}$ final state by using deep learning at a future
 high luminosity LHC (HL-LHC) run. With this method, a constraint of 
$-0.8<\lambda_3/\lambda_3^{\rm SM}<6.6$ at $66\%$ CL may be set with 3000 
fb$^{-1}$ of HL-LHC data. 
 Furthermore, for  HL-LHC runs, using $b\bar{b}\gamma \gamma $ channel for 
$HH$, the trilinear coupling can be constrained to $1.00<\lambda_3<6.22$
 at the $95$\% CL \cite{Chang:2019ncg}.   
Including various other decay 
channels, a precision range of $3.4-7.7$\% is proposed to be obtained 
for a $30$ ab$^{-1}$ integrated luminosity \cite{Mangano:2020sao}.  (See also
\cite{Park:2020yps}.)  

 Modification of the Higgs coupling or a dimension-six
operator in the SMEFT produces linear growth in energy of processes
involving vector bosons which are longitudinally polarized and Higgs in the final state~\cite{Henning:2018kys}. 
A  detailed analysis in the context of lepton and hadron colliders 
for processes such as $W_LW_L\rightarrow W_LW_LH$ and $W_L
W_L\rightarrow HHH$, is contained in Ref~\cite{Chen:2021rid} .

The current experimental bound from
di-Higgs production obtained by ATLAS on the ratio of Higgs self coupling
to its SM value is $-5.0 < \lambda_3/\lambda_3^{\mathrm SM}<12.0$  \cite{ATLAS:2019qdc} . 

For $e^{+}e^{-}$ colliders such as the FCC-ee \cite{FCC:2018evy}, ILC 
\cite{Bambade:2019fyw}, the CEPC \cite{CEPCStudyGroup:2018ghi}, and the
CLIC \cite{CLICdp:2018cto}, there have been several proposals for
using c.m. energies that cover a wide range starting from a few hundred GeV to a few TeV,  as well as for methods to determine
 trilinear Higgs coupling \cite{Azzurri:2021nmy, Blondel:2021ema}.

\section{\boldmath Loop contribution of the triple-Higgs coupling in $e^+e^- \to ZH$}
The process $e^+e^- \to ZH$ 
involves a $ZZ^*H$ vertex, which can be written as
\beq\label{ZZH}
\Gamma^{\mathrm ZZH}_{\mu\nu} = g_Zm_Z[(1+{\cal F}_1)g_{\mu\nu}
        + {\cal F}_2k_{1\nu}k_{2\mu}],
\eeq
where $k_1$, $k_2$ are the momenta (assumed directed inwards) 
of the gauge bosons carrying the respective
polarization indices $\mu$, $\nu$. This form assumes that the gauge
bosons
are either on-shell, satisfying $k_{i\mu}\epsilon^{\mu}(k_i) = 0$ 
($i=1,2$), or
couple to a conserved current, so that the terms with $k_{1\mu}$ or
$k_{2\nu}$ can be dropped. Here $m_Z$ is the $Z$ mass, and
$g_Z = g_W/\cos\theta_W$, $\theta_W$ being
the weak mixing angle. The quantities ${\cal F}_{1,2}$ are functions of
bilinear invariants constructed from the momenta.

Isolating the
contribution of the triple-Higgs coupling $\lambda_3$, the form factors
${\cal F}_{1,2}$ for the process  $Z^* \to ZH$ (we ignore 
the subprocess $W^{+*}W^{-*} \to H$, since the  measurement of the gauge
boson polarization requires decay into charged leptons) 
can be written at one-loop order in terms of the
Passarino-Veltman (PV) functions \cite{PV} as follows.
\beq\label{F1}
{\cal F}_1 ( k_1^2, k_2^2 ) = \frac{\lambda_3^{\rm SM}(1+ \kappa ) }{(4 \pi)^2} 
        \left( - 3B_0 -12 (m_Z^2 C_0 - C_{00}) - \frac{9 m_H^2}{2} (\kappa
        +1)B'_0 \right),
\eeq
\beq\label{F2}
{\cal F}_2 (k_1^2, k_2^2 ) = \frac{\lambda_3^{\rm SM}(1 + \kappa ) }{(4 \pi)^2}
        12(C_1 + C_{11} + C_{12}).
\eeq
For the process $Z^* \to ZH$, the arguments of the PV functions are
\beq\label{PVZH}
B_0 \equiv B_0 (m_H^2, m_H^2, m_H^2), \;\; C_0 \equiv
C_0(m_H^2,s,m_Z^2,m_H^2,m_H^2,m_Z^2),
\eeq
and analogously for the functions $B'_0$ and the tensor coefficients
$C_1, C_{11}$ and $C_{12}$.

The above expressions are to be evaluated to first order in the 
parameter
$\kappa$ for consistency, as there would be higher-loop contributions at
order $\kappa^2$ which are not being included. 
Use has been made of the package LoopTools \cite{looptools} to evaluate the PV integrals.

\section{\boldmath $Z$ polarization parameters and lepton angular
asymmetries}

\def\eehz{$e^+e^- \to ZH$~}

An earlier work \cite{Rao:2019hsp}
considered $Z$ polarization in \eehz in the presence of
anomalous $ZZH$ couplings, and discussed the role of angular
asymmetries of leptons produced in $Z$ decay. 
We use the formalism of that work in the presence of not only tree-level
anomalous coupling, but include the loop-induced couplings involving the
triple-Higgs coupling. 

We now use the notation of \cite{Rao:2019hsp}, rather than the one used
in eq. (\ref{ZZH}), for ease of comparison. The two notations are
related to each other in the way described below.
The vertex $Z_{\mu}(q)\rightarrow Z_{\nu}(k) H$  
in the process $e^+e^- \rightarrow Z^*(q) \rightarrow Z(k) H$
is written with the Lorentz structure
\begin{equation}\label{vertex}
 \Gamma^{ZZH}_{\mu \nu} =\frac{g}{\cos\theta_{W}}m_{Z} 
\left[ a_{Z}g_{\mu \nu}+
\frac{ b_{Z}}{m_{Z}^{2}}\left( q_{\nu}k_{\mu}-g_{\mu \nu}  q\cdot k\right) 
+\frac{\tilde b_{Z}}{m_{Z}^{2}}
\epsilon_{\mu \nu \alpha \beta} q^{\alpha} k^{\beta}\right]  
 \end{equation}
where $g$ is the $SU(2)_L$ coupling and $\theta_{W}$ is the weak mixing 
angle. The $a_{Z}$ and $b_{Z}$ terms are invariant under  CP, while  
the $\tilde b_{Z}$ term is  CP violating. In the SM, at 
tree level, the coupling $ a_{Z}=1$, whereas the other two couplings 
$b_{Z}$ and $\tilde b_{Z}$ vanish. We will now take $a_Z$ and $b_Z$ 
each to be a sum of a tree-level contribution from SMEFT 
 and a loop-level contribution from diagrams including 
an effective triple-Higgs coupling, and neglect the CP-violating
coupling $\tilde b_Z$. With this in mind, we are using a
modified notation, and now $a_Z$ and $b_Z$ will include also
contribution from ${\cal F}_1$ and ${\cal F}_2$ discussed above:
\begin{equation}
a_Z = a_Z^0 + {\cal F}_1 - (q\cdot k) {\cal F}_2,
\end{equation}
\begin{equation}
b_Z = b_Z^0 - m_Z^2 {\cal F}_2.
\end{equation}
In a low-energy SMEFT with a cut-off scale $\Lambda$, $a_Z$ and $b_Z$ 
would receive tree-level contributions, represented above by the
subscript 0, from the SM and from dimension-six 
terms like $\Phi^\dagger \Phi F_{\mu\nu}F^{\mu\nu} /\Lambda^2$,
where $\Phi$ is the scalar doublet field and
$F_{\mu\nu}$ is the field strength tensor.

The non-zero helicity amplitudes in the limit of massless initial 
states are \cite{Rao:2019hsp}
\begin{eqnarray}\label{helamp1}
M(-,+,+)&=&\frac{g^{2}m_{Z}\sqrt{{s}}(c_{V}+c_{A})}
{2\sqrt{2}\cos^{2}\theta_{W}({s}-m_{Z}^{2})}\left[a_Z- \frac{\sqrt{{s}}}{ m_{Z}^{2}}(E_{Z}b_{Z}+ i  \tilde b_{Z}\vert \vec{p}_{Z}\vert)
\right]\\ \nonumber
&& \times  (1-\cos\theta)\\\label{helamp2}
M(-,+,-)&=&\frac{g^{2}m_{Z}\sqrt{{s}}(c_{V}+c_{A})}
{2\sqrt{2}\cos^{2}\theta_{W}({s}-m_{Z}^{2})}\left[a_Z- \frac{\sqrt{{s}}}{ m_{Z}^{2}}(E_{Z}b_{Z}- i  \tilde b_{Z}\vert \vec{p}_{Z}\vert)
\right] \\ \nonumber
&& \times (1+\cos\theta)\\\label{helamp3}
M(-,+,0)&=&\frac{g^{2}E_Z\sqrt{{s}}(c_{V}+c_{A})}
{2\cos^{2}\theta_{W}({s}-m_{Z}^{2})}\left[a_Z-\frac{\sqrt{{s}}}{E_Z}
b_{Z}\right]  \sin\theta \\\label{helamp4}
M(+,-,+)&=&\frac{-g^{2}m_{Z}\sqrt{{s}}(c_{V}-c_{A})}
{2\sqrt{2}\cos^{2}\theta_{W}({s}-m_{Z}^{2})}\left[  a_Z-
  \frac{\sqrt{{s}}}{ m_{Z}^{2}}(E_{Z}b_{Z}+ i  \tilde b_{Z} \vert \vec{p}_{Z}\vert)\right] \\ \nonumber 
  && \times (1+\cos\theta)\\\label{helamp5}
M(+,-,-)&=&\frac{-g^{2}m_{Z}\sqrt{{s}}(c_{V}-c_{A})}
{2\sqrt{2}\cos^{2}\theta_{W}({s}-m_{Z}^{2})}\left[  a_Z-
  \frac{\sqrt{{s}}}{ m_{Z}^{2}}(E_{Z}b_{Z}- i  \tilde b_{Z}\vert \vec{p}_{Z}\vert)\right] \\ \nonumber
  && \times (1-\cos\theta)\\\label{helamp6}
M(+,-,0)&=&\frac{g^{2}E_Z\sqrt{{s}}(c_{V}-c_{A})}{
2\cos^{2}\theta_{W}({s}-m_{Z}^{2})}\left[a_Z-\frac{\sqrt{{s}}}{E_Z}
b_{Z}\right]  \sin\theta
\end{eqnarray}
Here the first two entries in $M$ denote the signs of the helicities of 
the $e^-$ and $e^+$, respectively, and the third entry is the 
$Z$ helicity. $\theta$ is the polar angle of the $Z$ relative to the
$e^-$ direction as the $z$ axis. $c_V$ and $c_A$ are respectively the
vector and axial-vector leptonic couplings of the $Z$, given by
\begin{equation}
c_V = \frac{1}{2}(-1+4\sin^2\theta_W),\;\;
c_A = -\frac{1}{2}.
\end{equation}

The density matrix elements for $e^-e^+\to ZH$ for the $Z$ spin summed
over the $e^+$ and $e^-$ helicities are 
derived from the helicity amplitudes, setting $\tilde b_Z =0$  
are given by
\begin{eqnarray}
\rho(\pm,\pm
)&=&\frac{g^{4}m^{2}_{Z}s}{8\cos^{4}\theta_{W}(s-m_{Z}^{2})^2}
\left[(c_{V}+c_{A})^{2}(1\mp\cos\theta)^{2}\right. \nonumber\\
&& \hskip -0.8cm \displaystyle 
\left. +(c_{V}-c_{A})^{2}(1\pm\cos\theta)^{2}\right]
\displaystyle \left| a_Z -  b_Z \frac{
E_{Z}\sqrt{s}}{m^{2}_{Z}}\right|^2  \\
\rho(0,0)&=&\frac{g^{4}E^{2}_{Z}s}{2\cos^{4}\theta_{W}(s-m_{Z}^{2})^2}\sin^{2}\theta
\, (c_V^2 + c_A^2) 
\left| a_Z  -b_{Z}
\frac{\sqrt{s}}{E_{Z}}\right|^2 \\
\rho(\pm,\mp
)&=&\frac{g^{4}m^{2}_{Z}s}{4\cos^{4}\theta_{W}(s-m_{Z}^{2})^2}\sin^{2}\theta
 \,(c_V^2 + c_A^2)
\left| a_Z -b_{Z}
\frac{
E_{Z}\sqrt{s}}{m^{2}_{Z}}\right|^{2} \\
\rho(\pm,0 )&=&\frac{g^{4}m_{Z}E_{Z}s}
{4\sqrt{2}\cos^{4}\theta_{W}(s-m_{Z}^{2})^2}\sin\theta
\nonumber \\
&& \hskip -.8cm 
\times \left[
(c_{V}+c_{A})^{2}(1\mp\cos\theta) 
 -(c_{V}-c_{A})^{2}(1\pm\cos\theta)\right] \nonumber\\
&& \hskip -.8cm 
\times \left[\vert a_Z \vert^2 
- a_Z b_Z^* \frac{ \sqrt{s}}{E_Z}
- a_Z^* b_Z \frac{ \sqrt{s}E_Z}{m_Z^2}
+ \frac{s}{m_Z^2} \vert b_Z \vert^2
\right]. 
\end{eqnarray}
Here $+$, $-$ and $0$ denote the $Z$ helicities.

We do not display the somewhat longer expressions for the density matrix
elements taking into account the polarizations $P_L$ and $\bar{P}_L$ of 
the 
electron and positron beams, respectively. However, the expressions are 
more 
compact on integration over $\cos\theta$, and these are displayed here.
We also include the appropriate phase space factor, so that the
expressions are normalized to give the correct total
cross section $\sigma$ as the trace of the density matrix:
\begin{equation}
\sigma = \sigma(+,+) + \sigma(-,-)+ \sigma(0,0).
\end{equation}
The integrated density matrix is given by
\begin{eqnarray}
 \sigma(\pm,\pm
)&=&\frac{2(1-P_L\bar P_L)g^{4}m^{2}_{Z}\vert \vec k_Z \vert}{192\pi
\sqrt{s} \cos^{4}\theta_{W}(s-m_{Z}^{2})^2}
(c_{V}^2+c_{A}^{2}-2P_L^{\rm eff}c_Vc_A) \nonumber\\
&& \hskip -0.8cm 
\times
\left| a_Z -b_{Z}
\frac{
E_{Z}\sqrt{s}}{m^{2}_{Z}}\right|^{2} \\
\sigma(0,0
)&=&\frac{2(1-P_L\bar P_L)g^{4}E^{2}_{Z}\vert\vec k_Z\vert}
{192\pi\sqrt{s}\cos^{4}\theta_{W}(s-m_{Z}^{2})^2}
(c_V^2 + c_A^2 -2P_L^{\rm eff}c_Vc_A)\nonumber \\ 
&& \hskip -0.8cm
\times \left| a_Z  -b_{Z}
\frac{\sqrt{s}}{E_{Z}}\right|^2 \\
\sigma(\pm,\mp)&=&\frac{(1-P_L\bar P_L)g^{4}
m^{2}_{Z}\vert \vec k_Z \vert }
{192\pi\sqrt{s}\cos^{4}\theta_{W}(s-m_{Z}^{2})^2}
 \,(c_V^2 + c_A^2 - 2P_L^{\rm eff}c_Vc_A)\nonumber \\ 
&& \hskip -0.8cm
\times \left| a_Z -b_{Z}
\frac{
E_{Z}\sqrt{s}}{m^{2}_{Z}}\right|^{2} \\
\sigma(\pm,0 )&=&\frac{(1-P_L\bar P_L)g^{4}m_{Z}E_{Z}\vert\vec k_Z\vert}
{256\sqrt{2}\sqrt{s}\cos^{4}\theta_{W}(s-m_{Z}^{2})^2}
 (2c_{V}c_{A} - P_L^{\rm eff}(c_V^2+c_A^2))
\nonumber\\
&& \hskip -.8cm 
\times \left[\vert a_Z \vert^2 
- a_Z b_Z^* \frac{ \sqrt{s}}{E_Z}
- a_Z^* b_Z \frac{ \sqrt{s}E_Z}{m_Z^2}
+ \frac{s}{m_Z^2} \vert b_Z \vert^2
\right]. 
\end{eqnarray}
In the above equations, $P_L^{\rm eff} = (P_L - \bar P_L)/ 
( 1 - P_L \bar P_L)$, and the indices $+$, $-$ and $0$ denote the $Z$
helicities.

It was shown that certain
angular asymmetries of the charged lepton produced in the decay of the
$Z$ can be simply related to vector and tensor
polarizations of the $Z$ \cite{
Boudjema:2009fz, Rahaman:2017qql, Rahaman:2016pqj, Rao:2019hsp} 
and hence to the spin density matrix of the
$Z$ in the process. Each of the asymmetries considered was found to be
dominated by one of the  anomalous couplings $\delta a_Z \equiv {\rm
Re}~a_Z -1$, ${\rm Re}~b_Z$ and
${\rm Im}~b_Z$ to linear order in the
anomalous couplings. Now that we include loop contributions of the
triple-Higgs couplings, these asymmetries could be measured 
experimentally
to put limits on linear combinations of the anomalous $ZZH$ 
couplings and the anomalous triple-Higgs coupling $\kappa$.  

Here we choose a particular angular asymmetry,  $A_{yz}$, 
the only one found to be proportional to the imaginary part of the 
anomalous coupling $b_Z$.  $A_{yz}$ is defined as 
\begin{equation}
A_{yz} 
\equiv\frac{\sigma(\cos\theta^{\ast}\sin\phi^{\ast}>0)
-\sigma(\cos\theta^{\ast}\sin\phi^{\ast}<0)}
{\sigma(\cos\theta^{\ast}\sin\phi^{\ast}>0)
+\sigma(\cos\theta^{\ast}\sin\phi^{\ast}<0)}
\end{equation}
and is related to the tensor polarization component
\begin{equation}
  T_{yz}=\frac{- i  \sqrt{3}\lbrace[\sigma(0,+)-\sigma(+,0)]-[\sigma(-,0)-\sigma(0,-)]\rbrace}{4\sigma} 
\end{equation}
by
\begin{equation}\label{asyz}
A_{yz}=\frac{2}{\pi}\sqrt{\frac{2}{3}} T_{yz}.
\end{equation}
Here, $\sigma(i,j)$ is the integral of $\rho(i,j)$ over the $Z$
azimuthal angle, $\sigma$ being the trace of $\sigma(i,j)$. 
The angles $\theta^\ast$ and $\phi^\ast$ are polar and azimuthal angles 
of the lepton in the rest frame of the $Z$. The $Z$ rest frame is 
reached by  a combination of boosts and rotations from the laboratory 
frame. In the laboratory frame, the $e^-$ momentum defines the positive 
$z$ axis, and the production plane of $Z$ is
defined as the $xz$ plane. While boosting to the $Z$ rest frame, the 
$xz$
plane is kept unchanged. Then, the angles $\theta^\ast$ and $\phi^\ast$
are  measured with respect to the would-be momentum of the $Z$.

Let us understand why the asymmetry $A_{yz}$ will be
proportional to Im~$b_Z$. For that we need to know the transformation
properties of $\cos\theta^\ast \sin\phi^\ast$ under CPT.
Under na\" ive time reversal T, all momenta change sign.
With the above definitions, it can be seen that under T,
the $Z$ momentum in the laboratory frame changes sign, as also the
momentum of the decay charged lepton, so the value of 
$\theta^\ast$ unchanged. On the other hand, since the normal to the $xz$
plane, which is along $\vec p_{e^-} \times \vec p_Z$, 
 does not change under T,  
the $y$ component of the decay-lepton momentum 
changes sign because $\vec p_{\ell^-}$ itself changes sign.
This makes the asymmetry $A_{yz}$ a T-odd asymmetry. At the same time,
the momenta of the $e^-e^+$ pair in the laboratory frame, 
as also those of the final charged
lepton pair in the $Z$ rest frame, are invariant under CP. The momentum
of the $Z$ does change sign under CP, and also
the normal to the $xz$ plane
changes direction. Therefore, $\cos\theta^\ast$ and $\sin\phi^\ast$ both
change sign. The combination 
$\cos\theta^\ast \sin \phi^\ast$ then remains invariant under CP. 
The asymmetry $A_{yz}$ is thus odd under CPT. 
As remarked earlier, for a CPT-odd quantity to get a
nonzero value, it should get contribution from an absorptive amplitude.
In our case, the only absorptive part which can interfere at linear
order with the SM contribution is 
from the imaginary part of $b_Z$. It can be checked that all other 
asymmetries are 
either even under CPT, or else odd under CP and so not possible at
one-loop level. 

Making use of expression for the density matrix elements 
derived in the presence of anomalous $ZZH$ couplings $a_Z$ and $b_Z$ 
(since no CP
violation is possible at one-loop level, we do not consider 
$\tilde b_Z$), we find that the asymmetry $A_{yz}$ is given by
\begin{equation}\label{asyzfull}
A_{yz}=\left(\frac{2c_V c_A-P_L^{\rm eff}(c_V^2+c_A^2)}{4(c_V^2 + c_A^2 
-2 P_L^{\rm eff}c_V c_A)}\right) 
\left(\frac{\vert \vec k_Z \vert^2
\sqrt{s}}{(E_Z^2+m_Z^2)m_Z}\right) 
\left(\frac{ {\rm Im}~(a_Z^{\ast}b_Z)}
{\vert a_Z \vert^2}\right).
\end{equation}

As can be seen, the asymmetry is proportional to Im~$b_Z$. Hence it
 will not get contributions
from the tree-level couplings, if they are assumed to be real. As
discussed earlier, in SM extensions like the 2HDM, $a_Z$ is real, though
it could be different from unity at tree level, and $b_Z=0$ at tree level. Im~$b_Z$
does get contribution at one-loop from the triple-Higgs coupling.
In fact, it can be written down as 
\begin{equation}
{\rm Im}~b_Z =  - m_Z^2 {\rm Im}~{\cal F}_2. 
\end{equation}
$A_{yz}$ can therefore be used to determine the triple-Higgs coupling
independent of the tree-level anomalous $ZZH$ couplings.
$a_Z$ also gets contribution from the triple-Higgs coupling at loop
level, but working to linear order in the anomalous couplings, Im $(
a_Z^{\ast} b_Z)/\vert a_Z \vert^2$ is simply Im $b_Z$. 

We now evaluate the contribution of the triple-Higgs coupling to the
asymmetry $A_{yz}$. The tree-level $b_Z$, if any, being real, will not
contribute to the asymmetry. The contribution of $a_Z$, appearing in
denominator of the asymmetry, will be restricted, for consistency 
to $a_Z = 1$, since the contribution of any anomalous coupling 
$\delta a_Z = a_Z - 1$ will appear only as $\delta a_Z ^2$, and hence
can be neglected. 

We include the possibility of longitudinal electron
polarization of 80\% and positron polarization of 30\% for the ILC,
and only polarized electrons with 80\% polarization for CLIC. 
There are studies which consider the possibility of incorporating
polarized beams at CEPC as well as FCC-ee, which we hope will allow
their advantage to be utilized.
Since we find the most advantageous configuration as
$(P_{e^-}, P_{e^+}) = ( -0.8, +0.3)$, we show results of 
 only this combination. Moreover, for simplicity and uniformity, we
assume that the full luminosity to be available for the polarized beam
combination, though many staging possibilities have been considered
while planning future experiments. 

That such a combination of beam polarization enhances the asymmetry can
be seen from the following arguments. As we argued earlier, $A_{yz}$
is odd under T. Now, all momenta transform in the same way under parity
P as under T, i.e., they reverse their signs. In addition, under P, the
helicities change sign. Thus, if P is a symmetry of the theory, $A_{yz}$
being odd under P would vanish. Our theory is not symmetric under P,
since the left-hand and right-handed couplings of $Z$ to leptons are
different. But
because $c_V \approx -0.06$ is numerically small, there is an 
approximate P
invariance, and the asymmetry turns out to be proportional to $c_V$, and
therefore small. In the presence of significant 
beam polarizations which are 
opposite in sign for the $e^-$ and $e^+$, the P symmetry is no longer an
approximate symmetry. There is quite a large P violation,  and as a
result, $A_{yz}$ is greatly enhanced.

We can see how this works out in practice from explicit expressions.
$A_{yz}$ arises from a combination of the imaginary parts of the 
density matrix elements $\rho(\pm,0)$. It can be seen from eqs.
(\ref{helamp1})-(\ref{helamp6})
that the contribution to these matrix elements from the $e^-$ and $e^+$
helicity combinations $(-,+)$ and $(+,-)$ occur with opposite signs,
in addition to the different couplings $(c_V+c_A)$ and $(c_V-c_A)$,
respectively. Since numerically $c_V$ is much smaller than $c_A$ in
magnitude, this results in a partial cancellation in the calculation of
$\rho(+,0)$ as well as that of $\rho(-,0)$, giving a coupling
dependence of $2c_Vc_A$ in the numerator of the asymmetry as compared 
to $c_V^2+c_A^2$ in the  
cross section appearing in the denominator of the asymmetry.
In the presence of polarization, the factors $(c_V+c_A)^2$ and 
$(c_V-c_A)^2$ get different polarization dependent factors, preventing
the partial cancellation, and thus enhancing the asymmetry.

For an asymmetry, the estimated error takes the form 
\begin{equation}\label{limit_A}
\delta A=\frac{\sqrt{1-A^{2}_{SM}}}{\sqrt{\sigma_{SM}\mathcal{L}}}
\end{equation}
with $\sigma_{SM}$ being the SM cross section for the process $e^+e^- 
\to Z^*H \to \ell\bar{\ell}H$ ($\ell= e, \mu$) at a collider with 
integrated luminosity $\mathcal{L}$ and
$A_{SM}$ is the corresponding value of asymmetry in the SM.

The efficiency of measurement of the cross section for $ZH$ production
with $Z$ decaying into lepton pairs is taken
to be 0.4\% for $\sqrt{s}=240$ GeV and luminosity 10 ab$^{-1}$
\cite{TLEP}. It is appropriately scaled for other luminosities.
Efficiencies for cross section measurement quoted in earlier works are 
0.9\% for ILC at 250 GeV and
luminosity 2 ab$^{-1}$ \cite{Asner,Fujii:2017vwa},
3.8\% at CLIC for $\sqrt{s}=350$~GeV and luminosity 500 fb$^{-1}$
\cite{Abramowicz:2016zbo},
and 0.5\% at CEPC for $\sqrt{s}=240$ or 250 GeV and luminosity 5.6 ab$^{-1}$
\cite{An:2018dwb}.

For our numerical calculations we make use of the following values of
parameters: $m_Z=91.1876$ GeV, $m_W=80.379$ GeV, $m_H=125.0$ GeV, 
$\sin^2\theta_W = 0.22$ and $G_F = 1.1663787\times 10^{-5}$ GeV$^{-2}$.

\section{Results}

The results are as follows.

We first list in Table \ref{imbz} values of Im~$b_Z$ which arise 
from the
one-loop triple-Higgs contribution for the value of $\kappa=1$ for
various values for c.m. energy.
\begin{table}
\centering
\begin{tabular}{|c|c|}
\hline
  $\sqrt{s}$& Im~$b_Z$\\
 (GeV) & (for $\kappa = 1$)\\
\hline
  240 &  $-3.62\times 10^{-4}$ \\
  250 &  $-4.91\times 10^{-4}$ \\
  350 &  $-8.22\times 10^{-4}$ \\
  365 &  $-8.09\times 10^{-4}$ \\
  380 &  $-7.93\times 10^{-4}$ \\
  500 &  $-6.13\times 10^{-4}$\\
\hline
\end{tabular}
\caption{Values of Im~$b_Z$ from one-loop triple-Higgs coupling
contributions for $\kappa=1$ and various values of $\sqrt{s}$}
\label{imbz}
\end{table}
We present in Table \ref{ayzlimits} the asymmetry $A_{yz}$ and the
limits that could be obtained using this asymmetry for several
colliders, with different energies and integrated luminosities. In cases
where polarized beams are likely to be available, we include the results
with unpolarized beams, as well with $e^-$ and $e^+$ polarizations of
$-0.8$ and $+0.3$ respectively. In case of CLIC, we give the result with
only electron beams polarized.

\begin{table}
\centering
\begin{tabular}{|c|c|c|c|c|c|c|}
\hline
Collider & c.m. & \multicolumn{2}{c|}{$10^4\times A_{yz}$} 
&Lumi-  & \multicolumn{2}{c|}{Limit }  \\
&energy&unpolarized &polarized  &nosity&unpolarized   &polarized  \\
&(GeV)&beams&beams&(ab$^{-1}$)&beams&beams\\
\hline
CEPC & 240 &$-0.159$&& 10 &  $506$ & \\
CEPC & 240 &$-0.159$&& 20 &  $358$ & \\
CLIC & 380 &$-2.88$&$-10.6$& 0.5 & $124$& $31.0 $  \\
FCC  & 240 &$-0.159$&& 10 &  $506$ & \\
FCC & 250 &$-0.314$&& 5 &   $362 $ & \\
FCC & 365 &$-2.64$&& 1.5 &   $78.2$ & \\
ILC & 250 &$-0.314$&$-1.23$& 2 &  $573 $ &  $119 $\\
ILC & 250 &$-0.314$&$-1.23$& 5 &   $362 $ & $75.3$\\
ILC & 350 &$-2.39$&$-9.38$& 30 & $19.4 $ &  $4.03 $\\
ILC & 500 &$-4.00$&$-15.7$& 4 &   $31.6 $ &  $6.57$\\
ILC & 500 &$-4.00$&$-15.7$& 10 &   $20.0 $ &  $4.16$\\
ILC & 500 &$-4.00$&$-15.7$& 30 & $11.5 $ &  $2.40 $\\
\hline
\end{tabular}
\caption{Values of the asymmetry $A_{yz}$ for $\kappa=1$ and 
1 $\sigma$ limits on $\kappa$ from $A_{yz}$ 
at various colliders with
different energies and luminosities. Values for CLIC are shown for 
unpolarized beams, as well as for $e^-$ beams polarized to $-80$\%. In
case of ILC, values are shown for unpolarized beams as well 
as with
$e^-$ and $e^+$ beam polarizations of $-0.8$ and $+0.3$, respectively.}
\label{ayzlimits}
\end{table}

The limit on $\kappa$ from
$A_{yz}$ with the same ILC parameters as mentioned, {\it viz.},
$\sqrt{s}=250$ GeV and ${\cal L}=2$ ab$^{-1}$, and using the same
sensitivity for $Z$ measurement as obtained from the literature,
comes out to be about
119. However, this limit has the advantage that it 
is strictly independent
of the tree-level contribution to the $ZZH$ coupling. 
To maintain this advantage and get a better limit,  
both energy and luminosity need to be pushed up. Thus,
 for $\sqrt{s}=500$ GeV, it is 4.16  
for ${\cal L}=10$ ab$^{-1}$ and 2.40 for ${\cal L}=30$ ab$^{-1}$. 

\begin{table}
\centering
\begin{tabular}{|c|c|c|c|c|c|c|}
\hline
Collider & c.m. & \multicolumn{2}{c|}{Cross section (fb) } 
&Lumi-  & \multicolumn{2}{c|}{Limit (\%)
}  \\
&energy& unpolarized&polarized &nosity& unpolarized  & polarized \\
&(GeV)&beams &beams&(ab$^{-1}$)&beams&beams\\
\hline
CEPC & 240 &247&& 10 &  $27.3$ & \\
CEPC & 240 &247&& 20 &  $19.3$ & \\
CLIC & 380 &107&127& $0.5$ & $1334$ & $1223$  \\
FCC & 240 &247&& 10 &  $27.3$ & \\
FCC & 250 &246&& 5 &   $44.3$ & \\
FCC & 365 &117&& 1.5 &   $506$ & \\
ILC & 250 &246&368& 2 &  $70.0$ &  $57.1$\\
ILC & 250 &246&368& 5 &  $44.3$ &  $36.1$\\
ILC & 350 &129&194& 30 & $81.3$ &  $66.4$\\
ILC & 500 &56.8&85.2& 4 &   $311$ & $254$\\
ILC & 500 &56.8&85.2& 10 &   $197$ &$161 $\\
ILC & 500 &56.8&85.2& 30 & $114$ &  $92.8$\\
\hline
\end{tabular}
\caption{Cross sections (in fb) for $\kappa=1$ and 
1 $\sigma$ limits (in per cent) on $\kappa$ from the cross section 
at various colliders with
different energies and luminosities, assuming no anomalous tree-level 
$ZZH$ contribution. Values for CLIC are shown for 
unpolarized beams, as well as $e^-$ beams polarized to $-80$\%. In
case of ILC, values are shown for unpolarized beams as well 
as with
$e^-$ and $e^+$ beams polarizations of $-0.8$ and $+0.3$, respectively.}
\label{cslimits}
\end{table}

While our main interest is to use measurements which are independent of
the tree-level anomalous couplings, for the sake of completeness, 
we also list in Table \ref{cslimits} the limits (in percent) 
that can be obtained on $\kappa$ from the measurement of the cross 
section, under the assumption
that there are no tree-level anomalous contributions, i.e., $\delta
a_{z} = 0$ and $b_Z = 0$. Table \ref{smcs} lists the SM cross sections
at various c.m. energies, including the contribution of the one-loop
triple-Higgs contribution.
\begin{table}
\centering
\begin{tabular}{|c|c|c|}
\hline
c.m. energy & \multicolumn{2}{c|}{Cross section (fb)} \\
(GeV) & unpolarized & polarized \\
\hline
240 & 243 & 365 \\
250 & 242 & 363\\
340 & 138 & 207\\
350 & 129 & 194\\
365 & 117 & 176\\
380 & 107 & 127\\
500 & 56.9 & 85.4 \\
\hline
\end{tabular}
\caption{Values of the SM cross sections
at various c.m. energies, including the contribution of the one-loop
triple-Higgs contribution, for unpolarized and polarized beams. 
In the polarized case $e^-$ and $e^+$ polarizations are respectively
$-0.8$ and $+0.3$, except in the case of the for 380 GeV,
where the $e^+$ polarization is zero.
 }
\label{smcs}
\end{table}

\section{Conclusions}
We have investigated the possibility of using a tensor polarization 
variable $T_{yz}$ of
the $Z$ in the process \eehz and a decay-lepton asymmetry $A_{yz}$
related to this variable to constrain the loop-level contribution to the
anomalous triple-Higgs coupling $\kappa$, independent of tree-level
anomalous $ZZH$ coupling. $A_{yz}$, an asymmetry dependent on the lepton
azimuthal angle in a specific frame, is found to be odd under na\" ive
time reversal operation, and is therefore proportional to the absorptive
part of the amplitude. It therefore isolates the loop-level
contributions. In case the triple-Higgs contribution is the dominant
one, we calculate the asymmetry and the  possible limit on $\kappa$ for
several collider energies and luminosities.

It is seen that a considerably high luminosity of at least 30 fb$^{-1}$
 is needed to constrain anomalous triple-Higgs coupling even at the
level of 200\% to 400\%. The better limit of about 240\% requires 
a c.m. energy of 500 GeV. 
The use of longitudinally polarized beams is also seen as absolutely 
essential to reach these sensitivities. 
Longitudinal beam $e^-$ polarization of $-80$\% and $e^+$ polarization
of $+30$\% improve the limits by a factor between 4 and 5.

While the limit which could be obtained by measuring the cross section
for the same combination of experimental parameters is much better,
about 93\%, this is possible only if the tree-level $ZZH$ contribution
is known to be negligible.

Note that we have used the formalism of \cite{Boudjema:2009fz}, which
entails a particular choice of frame of reference, as also
the charged-lepton decay channels for the $Z$. It is possible that by 
employing some different frame, as also including hadronic decay 
channels of the $Z$, the sensitivity may be improved. This investigation
would require detailed numerical simulations, which we do not attempt in
this work.

\noindent{
\bf Acknowledgement
} SDR acknowledges support from the Indian National Academy, New Delhi,
under the Senior Scientist programme.

\end{document}